\documentclass[12pt]{iopart}

\usepackage{amsfonts}
\usepackage{graphicx,color}

\newcommand{\tilt}{\tilde{t}}
\newcommand{\tilth}{\tilde{\theta}}
\newcommand{\tilph}{\tilde{\phi}}
\newcommand{\tilr}{\tilde{r}}

\begin{document}

\title{Zero, Normal and Super-radiant  Modes for Scalar and Spinor Fields
in Kerr-anti de Sitter Spacetime}

\author{Masakatsu Kenmoku$^1$, Yongmin Cho$^2$, Kazuyasu Shigemoto$^3$, Jong Hyuk Yoon$^4$}
\address{$^1$ Nara Science Academy, 178-2 Takabatake, Nara 630-8301, Japan}
\address{$^2$ Administration Building 310-4, Konkuk University, Seoul 05029, Korea}
\address{$^3$ Tezukayama University, 7-1-1 Tezukayama, Nara 631-8501, Japan}
\address{$^4$ School of Physics, Konkuk University, Seoul 05029, Korea}
\ead{$^1$m.kenmoku@cc.nara-wu.ac.jp, $^2$ymcho7@konkuk.ac.kr, $^3$shigemot@tezukayama-u.ac.jp, $^4$yoonjh@konkuk.ac.kr}
\vspace{0.1in}

\begin{indented}
\item[]August 2016
\end{indented}

\begin{abstract}

Zero and normal modes for scalar and spinor fields 
in Kerr-anti de Sitter spacetime 
are studied as bound state problem 
with Dirichlet and Neumann boundary conditions. 
Zero mode is defined as the momentum near the horizon to be zero: 
$p_{\rm H}=\omega-\Omega_{\rm H}m=0$, 
and is shown not to exist as physical state for both scalar and spinor fields.  
Physical normal modes satisfy the spectrum condition $p_{\rm H}>0$ as a result of 
non-existence of zero mode and the analyticity 
with respect to rotation parameter $a$ of Kerr-anti de Sitter black hole. 
Comments on the super-radiant modes and the thermodynamics of black hole are given 
in relation to the spectrum condition for normal modes.  
Preliminary numerical analysis on normal modes is presented.   

\end{abstract}

%%%%%%%%%%%%%%%%%%%%%%%%%%%%%%%%%%
\section{Introduction} 
\renewcommand{\theequation}{\thesection.\arabic{equation}}
\setcounter{equation}{0}

The interactions of black holes with matter fields 
are fundamental and important in observational astrophysics.
It is well-known that super-radiant phenomenon 
(outgoing intensity of matter fields becomes greater than ingoing intensity) 
can occur in rotating black hole spacetime.   
The successive occurrence of super-radiant phenomena leads to instability, 
which is called  "black hole bomb".     
Especially Kerr-anti de Sitter (AdS) spacetime 
leads to successive super-radiant phenomena because 
it plays the role of the reflection mirror \cite{Press1972,Chandrasekhar1983,
Cardoso2004,Kodama2008}. 

Also the stability of BH is related to the BH thermodynamics
\cite{Hawking1971,Carter1973}, 
which should be defined in equilibrium states.         
G. 't Hoot introduced the brick wall model in order to study BH thermodynamics \cite{tHoot1985} by imposing Dirichlet boundary condition on the event horizon of BH. The boundary condition defines normal modes 
(i.e., bound states) of matter fields 
and the sum of normal modes determines the partition function and the entropy of BH. 
The Boltzmann factor of the matter fields in the brick wall model becomes 
ill-defined, if the super-radiance for rotating BH occurs    
\cite{Mukohyama2000}.   

The super-radiant phenomenon is also important relating to the recent works on  
the star motion \cite{Cardoso2015,Rosa2015} and the radiation from axion 
\cite{Yoshino2014,Yoshino2015}. 
Here we list up important features of the super-radiant phenomena 
in rotating BH: 
\begin{enumerate}
\item[A.] 
The general condition of super-radiant phenomena for scalar fields is 
\begin{eqnarray}
\omega p_{\rm H}=
\omega(\omega-\Omega_{\rm H}m)<0 \ \ {\mbox{or}}\ \ 
1-{\Omega_{\rm H}m}/{\omega}<0,
\end{eqnarray}
where $\omega$, $p_{\rm H}$, 
$m$ and $\Omega_{\rm H}$ (defined in sec. 2.3) denote the frequency, 
momentum near the horizon, 
azimuthal angular momentum for scalar fields and angular velocity of BH, respectively 
\cite{Bekenstein1973,Unruh1973,Mano1997}.   
Usually the super-radiant condition is  
$p_{\rm H}=
\omega-\Omega_{\rm H}m<0$ under the implicit understanding 
that the frequency is positive $\omega>0$. 
The super-radiance occurring under this condition is classified as type 1 \cite{Kenmoku2015}. 
\item[B.]
The super-radiant phenomena for massless neutrinos and massive spinor fields 
do not occur, because 
the momenta near horizon is treated as positive definite 
$p_{\rm H}>0$ due to the masslessness of neutrinos as well as the positive frequency 
condition $\omega>0$ \cite{Unruh1973,Wagh1985,Maeda1976}. 
\item[C.]
We studied the super-radiant phenomena of scalar and spinor fields as scattering problem in Kerr spacetime using the Bargmenn-Wigner formulation \cite{Kenmoku2015}. 
We derived the consistency condition for the current conservation relation 
among scalar and spinor fields, in accordance with the general super-radiant condition in equation (1.1).
\begin{itemize}
\item[C1.]  The super-radiance with 
\begin{equation} 
\omega>0 , \ \ \ p_{\rm H}=\omega-\Omega_{\rm H}m<0 \ 
(\mbox{in this case}\  m>0)
\end{equation}
does not occur for
neither scalar nor spinor fields, which is "type 1" in our classification. 
\item[C2.]   The super-radiance with 
\begin{equation} 
\omega<0 , \ \ p_{\rm H}=\omega-\Omega_{\rm H}m>0 \ 
(\mbox{in this case}\  m<0)
\end{equation}
does occur for both scalar and spinor fields, which is "type 2", in our classification. 
\end{itemize}   
\end{enumerate}

Based on our previous works \cite{Kenmoku2015,Kenmoku2008}, 
we study in this paper the bound state as normal modes for scalar and spinor 
fields in four dimensional Kerr-AdS spacetime  and derive 
the physical spectrum condition for normal modes: $p_{\rm H}=\omega-\Omega_{\rm H}m>0$. 
Key points are non-existence of zero modes $(p_{\rm H}=0)$ and 
the analyticity of matter fields with respect to rotation parameter $a$ of BH.

The organization of this paper is as follows. 
In section 2, zero and normal modes are studied 
for scalar fields in Kerr-AdS spacetime. The spectrum condition for normal modes is derived. 
We shall show that the partition function in the brick wall model 
is well-defined and the super-radiant modes is type 2. 
In section 3, zero and normal modes are studied for spinor fields in Kerr-AdS 
spacetime. The spectrum condition for spinor fields is also derived, 
which turn out to be the same as that for scalar fields.  
In section 4, numerical analysis of normal modes for scalar fields 
in case of $\omega=\bar{\mu}$ (mass of scalar fields) will be presented. 
Summary is given in section 5.

%%%%%%%%%%%%%%%%%%%%%%%%%%%%%%%%%%%%%%%%%%%%%%%%%%%
\section{Scalar Fields in Kerr-AdS spacetime} 
\renewcommand{\theequation}{\thesection.\arabic{equation}}
\setcounter{equation}{0}
In this section we study zero mode and normal modes as bound state   
problem for scalar fields in four-dimensional Kerr-AdS spacetime. 
Let us consider the Kerr-AdS metric in Boyer-Lindquist coordinates:    
\begin{eqnarray}
ds^2=
&-&\frac{\Delta_{r}}{\rho^2}
\left(dt-\frac{a \sin^2{\theta}}{\Xi}d\phi\right)^2
+\frac{\Delta_{\theta}\sin^2{\theta}}{\rho^2}
\left(adt-\frac{r^2+a^2}{\Xi}d\phi\right)^2\nonumber\\
&+&\frac{\rho^2}{\Delta_{r}}dr^2+\frac{\rho^2}{\Delta_{\theta}}d\theta^2
\ , \\
\Delta_{r}&=&(r^2+a^2)(1+r^2\ell^{-2})-2Mr \ , \ 
\Delta_{\theta}=1-a^2\ell^{-2}\cos^2{\theta}\ ,\vspace{1em} \nonumber \\
\rho^2&=&r^2+a^2\cos^2{\theta}\ \ \ , \ \ \  
\Xi=1-a^2\ell^{-2}\ , \nonumber
\end{eqnarray}
where $M$, $a=J/M$, and $\ell=\sqrt{-3/{\Lambda}}$ denote the 
mass, rotation parameters of BH, and cosmological parameter, respectively. 
The event horizon $r_{\rm H}
$ is defined as the outer zero of factor $\Delta_{r}$.

The explicit expression of metrics $g_{\mu\nu}$ are obtained 
from the line element in eq. (2.1) given in appendix A.  

%%%%%%%%%%%%%%%%%%%%%%%%%%%%%%%%%%%%%%%%%%%% 
\subsection{Scalar field equations} 
The Lagrangian 
and field equations for complex scalar fields $\Phi(x)$ in Kerr-AdS spacetime is given by:
\begin{eqnarray}
\mathcal{L}=g^{\mu\nu}\partial_{\mu}\Phi^*(x)\partial_{\nu}\Phi(x)
-\bar{\mu}^2\Phi^*(x)\Phi(x) , \\
(\frac{1}{\sqrt{-g}}\partial_{\mu}\sqrt{-g}g^{\mu\nu}\partial_{\nu}
-\bar{\mu}^2)\Phi(x)=0,
\end{eqnarray}
where $\bar{\mu}$ denotes the mass of scalar fields.   
If we write $\Phi(x)$ as   
\begin{eqnarray}
\Phi 
={\rm e}^{-i\omega t}{\rm e}^{i m \phi}S(\theta) R(r), 
\end{eqnarray}
then the scalar field equations are separated into angular and radial field equations 
respectively:
\begin{eqnarray}
\left(
\frac{
\partial_{\theta}\sin{\theta}\Delta_{\theta}\partial_{\theta}}{\sin{\theta}}
-\frac{(a\omega\sin{\theta}-\Xi m/\sin{\theta})^2}{\Delta_{\theta}}
-\bar{\mu}^2a^2\cos^2{\theta}+\lambda \right)S(\theta)&=&0 
\\
\left(
\partial_{r}\Delta_{r}\partial_{r} 
+\frac{((r^2+a^2)\omega-\Xi am)^2}{\Delta_{r}}
-\bar{\mu}^2r^2 -\lambda
\right)R(r)&=&0 , 
\end{eqnarray}
where $\omega, m$, and $\lambda$ denote the frequency, azimuthal angular momentum 
of scalar fields, and the separation parameter, respectively. 

%%%%%%%%%%%%%%%%%%%%%%%%%%%%%%%%%%%%%%%%%%%%%%%%%%%%%%%%%
\subsection{ Normal modes and mode expansions }

Now we study the normal modes as bound states 
with Dirichlet or Neumann boundary conditions on the horizon:  
\begin{eqnarray}
R(r) = 0 \ \ \  \mbox{or} \ \ \partial_{r}R(r)=0 \ \ \ {\rm on} \ \ \ r=r_{\rm H},
\end{eqnarray}
with asymptotic condition:
\begin{equation}
R(r)\rightarrow 0 \ \ \ {\rm as}\ \ \  r\rightarrow \infty.
\end{equation}
\footnote{Note that we introduce a small positive constant $\epsilon$ in eq. (2.7) 
 as $r_{+}+\epsilon$ for regularization in case of necessity.}  
The following two identities can be obtained from eqs. (2.5) and (2.6):
\begin{eqnarray}
&&\int_{0}^{\pi} d\theta\sin{\theta}
\left((\omega^{*}-\omega')
\frac{
a^2\sin^2{\theta}(\omega^{*}+\omega')-2a\Xi m}{\Delta_{\theta}}
-(\lambda^{*}-\lambda')\right)\nonumber\\
&&\hspace{4em}\times
S_{\omega,m,\lambda}^{*}(\theta)S_{\omega',m,\lambda'}(\theta)
=0, 
\end{eqnarray}
for angular part and 
\begin{eqnarray}
&&\int_{r_{\rm H}}^{\infty} dr \left((\omega^{*}-\omega')\frac{(r^2+a^2)^2(\omega^{*}+\omega')-2(r^2+a^2)
\Xi am}{\Delta_{r}}-(\lambda^{*}-\lambda')
\right)
\nonumber\\
&&\hspace{2em}\times
R_{\omega,m,\lambda}^{*}(r)R_{\omega',m,\lambda'}(r)=0, 
\end{eqnarray}
for radial part. 

Let us define $X_{(\omega,\omega',\lambda,\lambda')}$ as follows
\begin{eqnarray}
&&X_{(\omega,\omega',\lambda,\lambda')}:=
{\displaystyle \int_{r_{\rm H}}^{\infty}} dr \int_{0}^{\pi} d\theta \nonumber\\
&&\times \frac{\sqrt{-g}}{\rho^2}
\left(
\left( 
-\frac{a^2\sin^2{\theta}}{\Delta_{\theta}}
+\frac{(r^2+a^2)^2}{\Delta_{r}}\right)
(\omega+\omega')
+\left(\frac{1}{\Delta_{\theta}}-\frac{r^2+a^2}{\Delta_{r}}
\right)2\Xi am\right)\nonumber\\
&&\times S_{\omega,m,\lambda}^{*}(\theta)S_{\omega',m,\lambda'}(\theta)
R_{\omega,m,\lambda}^{*}(r)R_{\omega',m,\lambda'}(r). 
\end{eqnarray}
Then from eqs. (2.9) and (2.10) we can obtain the following equations:
\begin{eqnarray}
(\omega^{*}-{\omega}')
X_{(\omega,{\omega}',\lambda,{\lambda}')}&=&0, \\ 
(\lambda ^{*}-{\lambda}')
X_{(\omega,{\omega}',\lambda,{\lambda}')}&=&0, 
\end{eqnarray}
which reduce to the real values of $\omega, \lambda$ and the orthonormal relations: 
\begin{eqnarray}
X_{(\omega,\omega',\lambda,\lambda')}
=\delta_{\omega,\omega'}\delta_{\lambda,\lambda'}.
\end{eqnarray} 
Let us define the inner product notation of $A$ and $B$ as follows   
\begin{eqnarray}
<A\ ,\ B>:=
\int_{\Sigma}d^3x 
\sqrt{-g}(-ig^{t\nu})
\left(A^{*}(t,x)\partial_{\nu}B(t,x)
-\partial_{\nu}A^{*}(t,x)B(t,x)\right)\ .\nonumber
\end{eqnarray}
The orthonormal relations are expressed: 
\begin{eqnarray}
<f_{\alpha}\ ,\ f_{\alpha'}>&=&-<f_{\alpha}^{*}\ ,\ f_{\alpha'}^{*}>
=\delta_{\alpha,\alpha'}^{(3)}\nonumber \\
<f_{\alpha}^{*}\ ,\ f_{\alpha'}>&=&<f_{\alpha}\ ,\ f_{\alpha'}^{*}>
=0 \ , 
\end{eqnarray}
with the eigenfunctions   
\[
f_{\omega,m,\lambda}:=\frac{1}{\sqrt{2\pi}}
{\rm e}^{-i\omega t}{\rm e}^{im\phi}
S_{\omega,m,\lambda}(\theta)R_{\omega,m,\lambda}(r)\ .
\]
where $\alpha=(\omega,m,\lambda)$ and $\alpha'=(\omega',m',\lambda')$ 
denote set of quantum numbers.

Let us expand the scalar field $\Phi$ and its conjugate momentum $\Pi$ 
in eigenfunctions $f_{\alpha}, f_{\alpha}^*$ as follows 
\begin{eqnarray}
\Phi(t,x)&=&\sum_{\alpha}\left(
a_{\alpha}f_{\alpha}(t,x)+b_{\alpha}^{\dagger}f_{\alpha}^{*}(t,x)
\right)
\ , \nonumber \\
\Pi(t,x)&=&\frac{\partial \mathcal{L}}{\partial \partial_{t}\Phi}
=-i\sum_{\alpha}(g^{tt}\omega-g^{t\phi}m)
(a_{\alpha}^{\dagger}f_{\alpha}^{*}(t,x)-b_{\alpha}f_{\alpha}(t,x)). 
\end{eqnarray}
The inverse relations can be found to  
\begin{eqnarray}
a_{\alpha}
&=& \int_{\Sigma} d^{3}x \sqrt{-g} \left(
i f_{\alpha}^{*}(t,x)\Pi^{\dagger}(t,x) 
-(g^{tt}\omega-g^{t\phi}m)f_{\alpha}^{*}(t,x)\Phi(t,x)
\right) \ ,\nonumber\\
b_{\alpha}^{\dagger}
&=&- \int_{\Sigma} d^{3}x \sqrt{-g} \left(
i f_{\alpha}(t,x)\Pi^{\dagger}(t,x)  
+(g^{tt}\omega-g^{t\phi}m)f_{\alpha}(t,x)\Phi(t,x) \right),
\end{eqnarray}
using the completeness relations:
\begin{equation}
\sum_{\alpha}(-ig^{t\nu})\left( f_{\alpha}^*(t,x)\partial_{\nu}f_{\alpha}(t,x') 
-\partial_{\nu}f_{\alpha}^*(t,x)f_{\alpha}(t,x')\right)
=\frac{1}{\sqrt{-g}}\delta^{(3)}(x-x').
\end{equation}  
Note that any states, including quasi-normal modes and super-radiant modes, can express 
in terms of eigenfunctions due to the completeness relation (2.18).

Canonical quantization relations are given by     
\begin{eqnarray}
\left[\Phi(t,x),\Pi(t,x')\right]
=\left[\Phi^{\dagger}(t,x),
\Pi^{\dagger}(t,x')\right]=\frac{i}{\sqrt{-g}}\delta^{(3)}(x-x'), 
\end{eqnarray}
which lead to the quantization of creation and annihilation operators:  
\[
[a_{\alpha},a_{\alpha'}^{\dagger}]
=[b_{\alpha},b_{\alpha'}^{\dagger}]
=\delta^{(3)}_{\alpha ,\alpha'}\ . 
\]

%%%%%%%%%%%%%%%%%%%%%%%%%%%%%%%%%%%%%%%%%%%%%%%%% 
\subsection{Conservation of energy and angular momentum}
Energy and angular momentum of scalar fields are given by  
\begin{eqnarray}
E&=& \int_{\Sigma} d^3x\sqrt{-g}(
-g^{tt}\partial_{t}\Phi^{\dagger}\partial_{t}\Phi
+g^{\phi\phi}\partial_{\phi}\Phi^{\dagger}\partial_{\phi}\Phi
\nonumber\\
&&+g^{rr}\partial_{r}\Phi^{\dagger}\partial_{r}\Phi 
+g^{\theta\theta}\partial_{\theta}\Phi^{\dagger}\partial_{\theta}\Phi)
\nonumber\\
L&=& \int_{\Sigma} d^3x
\sqrt{-g}(g^{tt}(\partial_{t}\Phi^{\dagger}\partial_{\phi}
\Phi
+\partial_{\phi}\Phi^{\dagger}\partial_{t }\Phi)
+2g^{t\phi}\partial_{\phi}\Phi^{\dagger}\partial_{\phi}\Phi),
\end{eqnarray}
which can be also written as  
\begin{eqnarray}
E=\sum_{\alpha}\omega(a_{\alpha}^{\dagger}a_{\alpha}
+b_{\alpha}b_{\alpha}^{\dagger})\ \ \ , \ \ \ 
L=\sum_{\alpha}m(a_{\alpha}^{\dagger}a_{\alpha}
+b_{\alpha}b_{\alpha}^{\dagger})\ . \nonumber
\end{eqnarray}
The effective energy of scalar field, as measured by the co-rotating observer, is given by  
\begin{eqnarray}
E-L\Omega_{\rm H}
= \sum_{\alpha}(\omega-\Omega_{\rm H}m)
(a_{\alpha}^{+}a_{\alpha}+b_{\alpha}b_{\alpha}^{\dagger}),
\end{eqnarray}
where 
\begin{equation}
\Omega_{\rm H}=a\Xi/(r_{\rm H}^2+a^2),
\end{equation}
denotes the angular velocity of BH 
on the horizon $r_{\rm H}$. 

Note that for normal modes $(\omega-\Omega_{\rm H}m\neq 0)$ near the horizon, 
we have ingoing and outgoing solutions 
with the momentum $p_{\rm H}$  with respect to a radial variable 
$r_{*}=\int dr (r^2+a^2)/\Delta_{r}$ as $\exp{(-ip_{\rm H}r_{*})}$ and $\exp{(ip_{\rm H}r_{*})}$.
The effective frequency $\omega_{\rm H}$ is the same in magnitude with the momentum 
$p_{\rm H}$ because scalar field is treated as massless near the horizon: 
\begin{equation}
\omega_{\rm H}=p_{\rm H}= \omega-\Omega_{\rm H}m.
\end{equation}

%%%%%%%%%%%%%%%%%%%%%%%%%%%%%%%%%%%%%%%%%%%%%%%%%%
\subsection{Non-existence of zero mode and spectrum condition for normal modes}
Now we consider the existence or non-existence of zero mode defined by  
$\omega-\Omega_{\rm H}m=0$. 
Radial equation for zero mode becomes:    
\begin{eqnarray}
\left(
\frac{d}{dr}\Delta_{r}\frac{d}{dr}+ \frac{(r^2-r_{\rm H}^2)^2\omega_{*}^2}{\Delta_r}
-\bar\mu^2r^2-\lambda \right)R_{\rm zero}(r)=0,
\end{eqnarray}
where $\omega_{*}=\Omega_{\rm H}m$ denotes 
the frequency of zero mode.  
The zero mode equation near horizon becomes from eq. (2.24):  
\begin{equation}
\left(\frac{d}{d r}(r-r_{\rm H})\frac{d}{dr} -{\alpha}\right)R_{\rm zero}(r)
\simeq 0,
\end{equation}
where $\alpha=(\bar{\mu}^2r_{\rm H}^2+\lambda)/X(r_{\rm H})$ 
with $X(r)=\partial_{r}\Delta_{r}$.  
General solution of eq. (2.24) is given by, using the Frobenius method: 
\begin{equation}
R_{\rm zero}(r) \simeq c_{1}(1+\alpha (r-r_{\rm H})) 
+c_{2}(\log (r-r_{\rm H})\, (1+\alpha  (r-r_{\rm H})) -2\alpha (r-r_{\rm H})), 
\end{equation}
where $c_{1}, c_{2}$ denote integration constants. 
If we impose the Dirichlet boundary condition, 
then from eq. (2.26), we find that $c_{1}=c_{2}=0$. 
Therefore non-trivial solution to eq. (2.25) 
which satisfies the Dirichlet boundary condition does not exist. 
We can also show the zero mode solution to eq. (2.25) does not satisfy 
the Neumann boundary condition either. 

%%%%%%%%%%%%%%%%%%%%%%%%%%%%%%%%%%%%%%%%%%%%%
From this result we can derive the spectrum condition of 
normal modes for scalar fields with Dirichlet or Neumann boundary conditions  
as follows:  
\begin{itemize}
\item[(i)]
When the specific rotation parameter $a=J/M=0$, 
the zero mode line is on $\omega=0$ and
the allowed physical modes are  $\omega>0$ in $\omega-m$ plane 
as in the standard scalar field theory in flat Minkowski spacetime. 
\item[(ii)]
When $a=J/M\neq 0$, the zero mode is defined by 
the line $\omega-\Omega_{\rm H}m=0$ in $\omega-m$ plane  
and the allowed physical modes should satisfy the spectrum condition:
\begin{eqnarray}
\omega-\Omega_{\rm H}m>0.
\end{eqnarray}
The reason is that any normal modes cannot across 
the zero mode line where the radial functions cannot  
satisfying the Dirichlet or Neumann boundary conditions. 
In this argument we take account of the analyticity with respect to rotation parameter $a$ 
under conditions that the outer horizon $r_{\rm H}$ should be well separated zero of 
$\Delta_{r}$ from other zeros and the rotation parameter should be less than the cosmological  parameter $a<\ell$ for regularity of $\Delta_{\theta}$ and $\Xi$. 
\end{itemize}

Let us comment on super-radiant modes and the brick wall model.   
\begin{itemize}
\item[(1)] 
We can apply the spectrum condition for normal modes to super-radiant modes 
because any modes can express in terms of eigenfunctions due to the completeness relation 
in eq. (2.18). 
Then allowed types of super-radiant modes taking account of general super-radiant 
condition given in equation (1.1) are the following two:  
\begin{enumerate}
\item[type 1]$ (\omega>0, \omega-\Omega_{\rm H}m<0)$  is unphysical and does not occur
because this type contradicts with the spectrum condition of (2.27).  
\item[type 2] $(\omega<0, \omega-\Omega_{\rm H}m)>0$ are physical 
which coincides with both equations (1.1) and (2.27).  
\end{enumerate}
In type 2 super-radiance, annihilation operators of particles with 
$\omega<0, m<0$ are understood as creation operators of antiparticles with 
$\omega>0, m>0$ as the interpretation in quantized field theory.  
Also, we mention that the results obtained here are consistent with 
our previous work basen on the Bargmann-Wigner formulation for scattering problem 
in Kerr spacetime \cite{Kenmoku2015}. 
\item[(2)]
The partition function in the Kerr-AdS black hole background spacetime 
becomes well-defined:
\begin{eqnarray}
Z={\rm Tr} \exp{(-\beta_{\rm H}(E-L\Omega_{\rm H}))}
\end{eqnarray}
considering the expression of effective energy (2.21) and 
the spectrum condition (2.27).  
\end{itemize}

%%%%%%%%%%%%%%%%%%%%%%%%%%%%%%%%%%%%%%%%%%%%%%%%%%%%%
%%%%%%%%%%%%%%%%%%%%%%%%%%%%%%%%%%%%%%%%%%%%%%%%%%%%%
\section{Spinor fields in Kerr-AdS spacetime}
\renewcommand{\theequation}{\thesection.\arabic{equation}}
\setcounter{equation}{0}
Next we consider zero and normal modes as bound states of massive spinor fields  
in Kerr-AdS spacetime.  
We study the spectrum condition and investigate the allowed type of 
super-radiant modes for spinor case, which can be compared with that for scaler case 
studied in the previous section.  

In order to study the Dirac equation in curved spacetime, 
the local tetrads $\{ \hat{e}_{t}, \hat{e}_{\theta}, \hat{e}_{\phi}, \hat{e}_{r},\}$ 
are introduced at each point of 
the general curved spacetime. The line element expressed in equation (2.1) 
is given as 
\begin{eqnarray}
ds^2&=& -{\hat{e}_{t}}^2+ \hat{e}_{\theta}^2+ \hat{e}_{\phi}^2+ \hat{e}_{r}^2,
\nonumber\\
\hat{e}_{t}&=&\frac{\sqrt{\Delta_{r}}}{\rho}(dt-\frac{a\sin^2{\theta}}{\Xi}d\phi), \ \ 
\hat{e}_{\theta}=\frac{\rho}{\sqrt{\Delta_{\theta}}}d\theta,  \nonumber\\
\hat{e}_{\phi}&=&\frac{\sqrt{\Delta_{\theta}}\sin{\theta}}{\rho}
(\frac{r^2+a^2}{\Xi}d\phi-adt), \ \ 
\hat{e}_{r}=\frac{\rho}{\sqrt{\Delta_{r}}}dr.
\end{eqnarray} 
The relation between the local tetrad and the general curved coordinate defines  
Viervein $b_{\mu}^{i}$ as $\hat{e}^{i}=b_{\mu}^{i}d{x}^{\mu}$,  
where the Greek letters ($\mu, \nu ...$) denote curved spacetime indices,  
and the Latin letters (i,j , ...) denote local tetrads.
The explicit expressions of Vierbeins 
for Kerr-AdS spacetime are given in appendix B.

%%%%%%%%%%%%%%%%%%%%%%%%%%%%%%%%%%%%%%%%%%
\subsection{The Dirac equation in Kerr spacetime} 
The Dirac equation in general curved spacetime is given by   
\begin{eqnarray}
(\gamma^{\mu}(\partial_{\mu}+\frac{1}{4}\omega^{ij}_{\mu}\gamma_{ij}) +{\bar  \mu} )\Psi(x)
=0
\end{eqnarray}
where 
$\omega^{ij}_{\mu}$, $\gamma_{ij}$, and $\bar\mu$ denote the spin connection,  
the anti-symmetric product of Dirac gamma matrices, and the mass of spinor fields, 
respectively.  
The spin connections are divided into two terms:   
\begin{eqnarray}
\gamma^{\mu}\frac{1}{4}\omega^{ij}_{\mu}\gamma_{ij}&=&
\frac{1}{2\sqrt{-g}}\gamma^{i}\partial_{\mu}(\sqrt{-g}b^{\mu}_{i})
+\frac{1}{4}b^{\mu}_{i}b^{\nu}_{j}\partial_{\mu}b_{\nu k}\gamma^{ijk},
\end{eqnarray}
whose derivation is in Appendix C. 
The first term of the right-hand side of eq. (3.3) is geometric in origin 
and the second term has spinorial origin. 
The totally anti-symmetric product of Dirac matrices is denoted by 
 $\gamma^{ijk}$.  
Each term is calculated as
\begin{eqnarray}
\frac{1}{2\sqrt{-g}}\gamma^{i}\partial_{\mu}(\sqrt{-g}b^{\mu}_{i})
&=&\frac{\gamma^{\tilr}}{2\rho}(\frac{r\sqrt{\Delta_{r}}}{\rho^2}
+\frac{\partial_{r}\Delta_{r}}{2\sqrt{\Delta_{r}}})\nonumber\\
&+&\frac{\gamma^{\tilth}}{2\rho}(-\frac{a^2\cos{\theta}\sin{\theta}}{\rho^2}\sqrt{\Delta_{\theta}}+\cot{\theta}\sqrt{\Delta_{\theta}}
+\frac{\partial_{\theta}\Delta_{\theta}}{2\Delta_{\theta}}), \\
\frac{1}{4}b^{\mu}_{i}b^{\nu}_{j}\partial_{\mu}b_{\nu k}\gamma^{ijk}&=&
\frac{a}{2\rho^3}(\gamma^{rt\phi}r\sin{\theta}\sqrt{\Delta_{\theta}}
-\gamma^{\theta t \phi} \cos{\theta}\sqrt{\Delta}_{r}), 
\end{eqnarray}

Then the explicit form of Dirac equation in Kerr-AdS spacetime in 
the Boyer-Lindquist coordinate becomes: 
\begin{eqnarray}
&&\Bigl(\frac{\gamma^{0}}{\rho\sqrt{\Delta_{r}}}{((r^2+a^2)\partial _{t}+a\Xi\partial_{\phi})}
+\frac{\gamma^{2}}{\rho\sqrt{\Delta_{\theta}}}
(a\sin{\theta}\partial_{t}+\frac{\Xi}{\sin{\theta}}\partial _{\phi})\nonumber\\
&+&\frac{\gamma^{3}\sqrt{\Delta_{r}}}{\rho}\partial_{r}
+\frac{\gamma^{1}\sqrt{\Delta_{\theta}}}{\rho}\partial_{\theta}\nonumber\\
&+&\frac{\gamma^{3}}{2\rho}(\frac{r\sqrt{\Delta_{r}}}{\rho^2}+\frac{\partial_{r}\Delta_{r}}{2\sqrt{\Delta_{r}}})
+\frac{\gamma^{1}}{2\rho}(-\frac{a^2\cos{\theta}\sin{\theta}}{\rho^2}\sqrt{\Delta_{\theta}}+\cot{\theta}\sqrt{\Delta_{\theta}}
+\frac{\partial_{\theta}\Delta_{\theta}}{2\Delta_{\theta}}) \nonumber\\
&+&\frac{a\gamma_{5}}{2\rho^3}(\gamma^{1}r\sin{\theta}\sqrt{\Delta_{\theta}}
+\gamma^{3} \cos{\theta}\sqrt{\Delta}_{r})+ {\bar\mu} \Bigr)\Psi(x)=0. 
\end{eqnarray}
The representation of Dirac matrices are:
\begin{eqnarray}
\gamma^{\tilt}&=&\gamma^{0}=-i\left(
\begin{array}{cc}
1 & 0\\
0 & -1
\end{array}
\right)
, \ \ \gamma^{\tilth}=\gamma^{1}=\left(
\begin{array}{cc}
0 & -i\sigma_{1}\\
i\sigma_{1} & 0
\end{array}
\right)
, \nonumber\\ 
\gamma^{\tilph}&=&\gamma^{2}=\left(
\begin{array}{cc}
0 & -i\sigma_{2}\\
i\sigma_{2} & 0
\end{array}
\right), \ \ 
\gamma^{\tilr}=\gamma^{3}=\left(
\begin{array}{cc}
0 & -i\sigma_{3}\\
i\sigma_{3} & 0
\end{array}
\right), \nonumber\\
\gamma_{5}&=&-i\gamma^{0}\gamma^{1}\gamma^{2}\gamma^{3}=
-\left(
\begin{array}{cc}
0 & I\\
I & 0
\end{array}
\right), 
\end{eqnarray} 
where $\sigma_{a}\ (a=1,2,3)$ are Pauli matrices.

With the ansatz 
\begin{eqnarray}
\Psi(x)=\frac{\exp{(-i\omega t +im\phi)}}{(\Delta_{r}\Delta_{\theta}\rho^2\sin^2{\theta})^{1/4 }}F(r, \theta), 
\end{eqnarray}
where  $\omega, m$ denote the frequency and azimuthal quantum number, respectively,   
the Dirac equation becomes 
\begin{eqnarray}
&&\Bigl(\frac{-i\gamma^{0}}{\rho\sqrt{\Delta_{r}}}{((r^2+a^2)\omega-a\Xi m)}
+\frac{-i\gamma^{2}}{\rho\sqrt{\Delta_{\theta}}}
(a\sin{\theta}\omega-\frac{\Xi m}{\sin{\theta}})\nonumber\\
&+&\frac{\gamma^{3}\sqrt{\Delta_{r}}}{\rho}\partial_{r}
+\frac{\gamma^{1}\sqrt{\Delta_{\theta}}}{\rho}\partial_{\theta}\nonumber\\
&+&\frac{a\gamma_{5}}{2\rho^3}(\gamma^{1}r\sin{\theta}\sqrt{\Delta_{\theta}}
+\gamma^{3} \cos{\theta}\sqrt{\Delta}_{r})+ {\bar\mu} \Bigr)F(r, \theta)=0, 
\end{eqnarray}
where the first term on the right-hand side of eq. (3.3) is eliminated.  
Furthermore, let us expand the spinor wave function $F(r, \theta)$ 
in chiral eigenfunctions $f_{\pm}$ as follows: 
\begin{eqnarray}
F(r, \theta)=(\frac{\chi^*}{\chi})^{1/4}f_{+}+(\frac{\chi}{\chi^*})^{1/4}f_{-} ,\ \ \ \gamma_{5}f_{\pm}=\pm f_{\pm},
\end{eqnarray}
with $\chi=r+ia\cos{\theta}$. Then the Dirac equation becomes 
\begin{eqnarray}
&&{\chi^*}^{1/2}
\Bigl(\frac{-i\gamma^{0}}{\rho\sqrt{\Delta_{r}}}{((r^2+a^2)\omega-a\Xi m)}
+\frac{-i\gamma^{2}}{\rho\sqrt{\Delta_{\theta}}}
(a\sin{\theta}\omega-\frac{\Xi m}{\sin{\theta}})\nonumber\\
&+&\frac{\gamma^{3}\sqrt{\Delta_{r}}}{\rho}\partial_{r}
+\frac{\gamma^{1}\sqrt{\Delta_{\theta}}}{\rho}\partial_{\theta}
+ {\bar\mu} \Bigr)f_{+}(r, \theta) \nonumber\\
&+&{\chi}^{1/2}
\Bigl(\frac{-i\gamma^{0}}{\rho\sqrt{\Delta_{r}}}{((r^2+a^2)\omega-a\Xi m)}
+\frac{-i\gamma^{2}}{\rho\sqrt{\Delta_{\theta}}}
(a\sin{\theta}\omega-\frac{\Xi m}{\sin{\theta}})\nonumber\\
&+&\frac{\gamma^{3}\sqrt{\Delta_{r}}}{\rho}\partial_{r}
+\frac{\gamma^{1}\sqrt{\Delta_{\theta}}}{\rho}\partial_{\theta}
+ {\bar\mu} \Bigr)f_{-}(r, \theta)=0,
\end{eqnarray}
where the second term on the right-hand side of eq. (3.3) is eliminated.
Let $f_{\pm}(r, \theta)$ be such that  
\begin{eqnarray}
f_{+}(r, \theta)&=&\frac{1}{\sqrt{2}}\left(
\begin{array}{c}
\zeta \\
-\zeta 
\end{array}
\right), \ \ \ 
f_{-}(r, \theta)
=\frac{1}{\sqrt{2}}\left(
\begin{array}{c}
\eta \\
\eta 
\end{array}
\right), \nonumber\\
\eta(r, \theta)&=&\left(
\begin{array}{c}
R_{1}(r)S_{1}(\theta) \\
R_{2}(r)S_{2}(\theta) 
\end{array}
\right),\ \ \ 
\zeta(r, \theta)=\left(
\begin{array}{c}
R_{2}(r)S_{1}(\theta) \\
R_{1}(r)S_{2}(\theta) 
\end{array}
\right). 
\end{eqnarray}  
Then the Dirac equation reduces to the following set of the ordinary differential equations: 
\begin{eqnarray} 
(\sqrt{\Delta_{\theta}}\partial_{\theta}+
\frac{a\omega\sin^2{\theta}-{\Xi m}}{\sqrt{\Delta_{\theta}}\sin{\theta}}) S_{1}(\theta)
=(-a{\bar\mu}\cos{\theta}+\kappa)S_{2}(\theta), \\
(\sqrt{\Delta_{\theta}}\partial_{\theta}-
\frac{a\omega\sin^2{\theta}-{\Xi m}}{\sqrt{\Delta_{\theta}}\sin{\theta}}) S_{2}(\theta)
=(-a{\bar\mu}\cos{\theta}-\kappa)S_{1}(\theta), \\
(\sqrt{\Delta_{r}}\partial_{r}-i\frac{(r^2+a^2)\omega-a\Xi m}{\sqrt{\Delta_{r}}})R_{1}(r)
=(\kappa-i{\bar\mu} r )R_{2}(r), \\
(\sqrt{\Delta_{r}}\partial_{r}+i\frac{(r^2+a^2)\omega-a\Xi m}{\sqrt{\Delta_{r}}})R_{2}(r)
=(\kappa+i{\bar\mu} r )R_{1}(r), 
\end{eqnarray}
where $\kappa$ is the separation parameter. 
The Dirac equations of the form (3.13)-(3.16) are consistent with 
 other previous works for Schwarzschild and Kerr spacetime 
\cite{Chandrasekhar1976,Belgiorno2010,Mao2011,Cebeci2012,Dolan2015}.

%%%%%%%%%%%%%%%%%%%%%%%%%%%%%%%%%%%%%%%%%%%%%%%%%%%%%%%%%%%%
\subsection{Non-existence of zero modes for spinor fields}
Now we study zero mode for spinor field.  
Radial equations for zero mode $(p_{\rm H}=\omega-\Omega_{\rm H}m=0)$ become 
\begin{eqnarray}
(\sqrt{\Delta_{r}}\partial_{r}-i\frac{(r^2-r_{\rm H}^2)\omega_{*}}{\sqrt{\Delta_{r}}})R_{1 \rm zero}(r)
=(\kappa-i{\bar\mu} r )R_{2 \rm{zero}}(r), \nonumber\\
(\sqrt{\Delta_{r}}\partial_{r}+i\frac{(r^2-r_{\rm H}^2)\omega_{*}}{\sqrt{\Delta_{r}}})R_{2 \rm zero}(r)
=(\kappa+i{\bar\mu} r )R_{1 \rm{zero}}(r), 
\end{eqnarray} 
where $\omega_{*}=\Omega_{\rm H}m=ma\Xi/(r_{\rm H}^2+a^2)$. 
Near the horizon, zero mode  equations (3.17) become 
\begin{eqnarray}
\sqrt{r-r_{\rm H}}\partial_{r}R_{1 \rm zero}(r)
\simeq \beta R_{2 \rm{zero}}(r),  \ \ 
\sqrt{r-r_{\rm H}}\partial_{r}R_{2 \rm zero}(r)
\simeq \beta^* R_{1 \rm{zero}}(r),
\end{eqnarray} 
where $\beta=(\kappa-i\bar{\mu}r_{\rm H})/\sqrt{X(r_{\rm H})}$ with 
$X(r)=\partial_{r}\Delta_{r}$.
General solutions to eq. (3.18) are given by 
\begin{eqnarray} 
R_{1 \rm zero}(r) &\simeq& d_{1}\exp{(2|\beta|\sqrt{r-r_{\rm H}})} 
+ d_{2}\exp{(-2|\beta|\sqrt{r-r_{\rm H}})}, \nonumber \\ 
R_{2 \rm zero}(r) &\simeq& (\beta^*/\beta)^{1/2}(d_{1}\exp{(2|\beta|\sqrt{r-r_{\rm H}})} 
- d_{2}\exp{(-2|\beta|\sqrt{r-r_{\rm H}})}),
\end{eqnarray} 
where $d_{1}$ and $d_{2}$ are integration constants. 

The Dirichlet and Neumann boundary conditions for spinor fields 
require the spacial component 
(in the present case, the radial component) of conserved current on the horizon to vanish: 
\begin{eqnarray}
J^{r}(x)=i\sqrt{-g}\bar{\Psi}(x)\gamma^{r}\Psi(x)=0\ \ \ {\mbox{on}}\ \ \ r=r_{\rm H} 
\end{eqnarray}
where $\bar{\Psi}=i\Psi^{\dagger}\gamma^{0}$. 
The explicit expression for $J^{r}$ is given by   
\begin{eqnarray}
J^{r}=\frac{1}{\sqrt{\Delta_{\theta}}\Xi}(|R_1|^2-|R_2|^2)(|S_{1}|^2+|S_2|^2),
\end{eqnarray} 
where we used eqs. (3.8), (3.10) and (3.12).
There are two solutions of $J^{r}=0$, 
\begin{equation} 
({\rm i}) \ \ \ R_1= (\beta/\beta^{*})^{1/2}R_2,\ \ \
({\rm ii}) \ \ \ R_1=-(\beta/\beta^{*})^{1/2}R_2,
\end{equation}
which correspond to the Dirichlet and Neumann boundary conditions, respectively. 
\footnote{
The Dirichlet and Neumann boundary conditions for spinor fields in eq. (3.22) are certified 
by the Bargmann-Wigner formulation too.}  
Here a constant phase factor $ (\beta/\beta^{*})^{1/2}$ is introduced, where 
 $\beta$ is in eq. (3.18).   
(Note that the additional phase factor reduces to unity in the 
massless limit: $\beta/\beta^{*}\rightarrow 1$ as $\bar{\mu}\rightarrow 0$). 
The zero mode solution in equation (3.19) satisfies neither (i) nor (ii) in eq. (3.22). 
This means that the zero mode of spinor fields does not exist as physical states.  
 
From the non-existence of zero mode for spinor fields 
we can derive the spectrum condition 
of normal modes for spinor fields satisfying $p_{\rm H}=\omega-\Omega_{\rm H}m>0$ 
as in the scalar field case.   

As a consequence of the spectrum condition, type 1 super-radiance 
($\omega>0, \omega-\Omega_{\rm H}m<0$) cannot occur whereas  
type 2 super-radiance ($\omega<0,  \omega-\Omega_{\rm H}m>0$) does occur 
for spinor fields as well as scalar fields.    
 
%%%%%%%%%%%%%%%%%%%%%%%%%%%%%%%%%%%%%%%%%%%%%%
\section{Numerical analysis on normal modes for scalar fields} 
\renewcommand{\theequation}{\thesection.\arabic{equation}}
\setcounter{equation}{0}

Now we present numerical analysis of normal modes for scalar fields 
in Kerr spacetime with the Dirichlet boundary condition and  
with special choice of parameter $\omega=\bar{\mu}$.  
In this case the separation parameter is set to the value  $\lambda=2(2+1)-2ma\omega+a^2\omega^2$. 
Parameters in the theory are chosen for the cosmological 
parameter and the black hole mass as $\Lambda=-3/\ell^2=0$ and $M=4$, 
and for the frequency, azimuthal quantum number of spinor fields 
as $\omega=\bar{\mu}=0.1645$, $m=0, \pm 2$. 
In Figure 1, the zero mode line, the ground state mode with respect to radial 
component denoted as $\omega_{0}$ and the first excited state modes 
as $\omega_{1}$ are shown.  
We can recognize that the normal modes with Dirichlet boundary condition on 
the horizon lie above the zero mode line and 
the spectrum condition is realized.

\begin{figure}[h]
\begin{center}
\includegraphics[height=5cm,width=8cm]{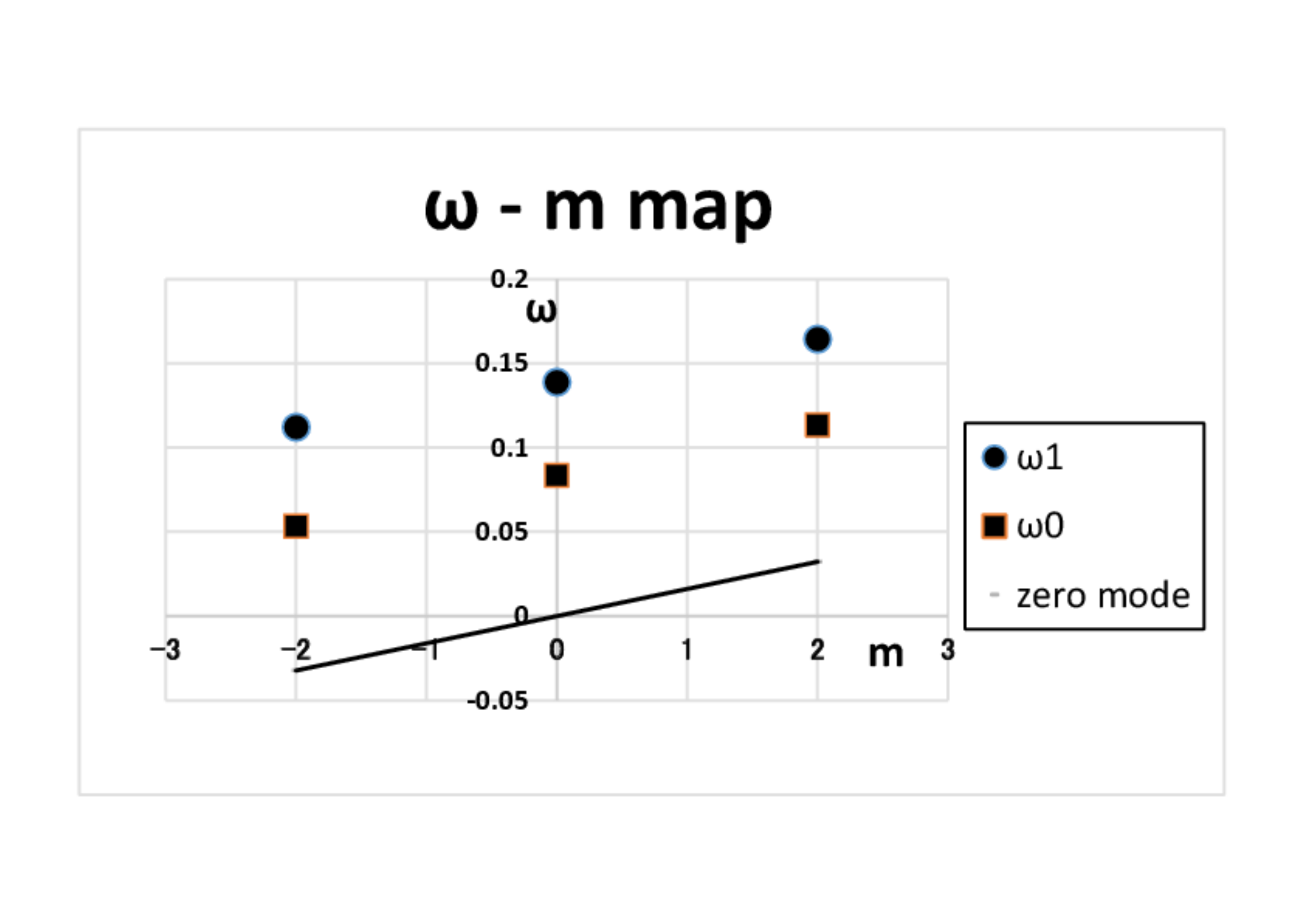} 
\caption{The plot of $\omega$ vs. m}
\label{figure:No1}
\end{center}
\end{figure}

%%%%%%%%%%%%%%%%%%%%%%%%%%%%%%%%%%%%%%%%%%%%

\section{Summary} 
\renewcommand{\theequation}{\thesection.\arabic{equation}}
\setcounter{equation}{0}
Zero and normal modes are studied for scalar and spinor fields 
in Kerr-AdS spacetime in case of the Dirichlet and  Neumann boundary conditions. 
\begin{itemize}
\item[S1]
For zero mode ($\omega-\Omega_{\rm H}m=0$),  
radial eigenfunctions that satisfy the Dirichlet or Neumann boundary condition 
does not exist for both scalar and spinor fields. 
\item[S2]
The spectrum condition for physical normal modes are 
in the range of $\omega-\Omega_{\rm H}m>0$, because 
(i) normal modes are in the positive omega region $\omega>0$ in 
no rotating cases $a=J/M=0$ and  
(ii) wave functions must be analytic with respect to the rotation parameter $a$. 
The spectrum condition holds for both scalar and spinor fields. 
\item[S3]
We apply the spectrum condition for normal modes (i.e., bound state problem)  
to the super-radiant phenomena, we find that the type 2 super-radiances  
($\omega<0 $ and $ \omega-\Omega_{\rm H}m>0$) are possible for both 
scalar and spinor fields.  
Note that the type 1 super-radiances $(\omega>0, \omega-\Omega_{\rm H}m<0)$ 
does not occur.  
\item[S4]
The preliminary numerical analysis with the special choice 
of parameters: $\omega=\bar{\mu}$ and $\Lambda=0$ 
 supports the previous theoretical analysis discussed in sections 2 and 3.  
\item[S5] 
The partition function in brick wall model is well-defined because of the spectrum condition 
$\omega-\Omega_{\rm H}m>0$.  
\end{itemize}

Note that the result on the super-radiance is consistent with 
our previous works 
where the  Bargmann-Wigner formulation was applied to scattering problem  
for both bosons and fermions in Kerr spacetime \cite{Kenmoku2015}, 
and normal modes for scalar fields in BTZ spacetime was studied analitically and numerically \cite{Kuwata2008}. 
Notice also that as to type 2 super-radiance,  
the annihilation operators of particles with negative frequency 
and negative azimuthal quantum number 
correspond to the creation operators of antiparticles 
with positive frequency and positive azimuthal 
quantum number as in standard quantum fields theory. 
This interpretation supports the idea proposed by Penrose \cite{Penrose1969}. 
Type 2 super-radiance is natural in the classical picture of super-radiant phenomena 
in rotating spacetime.       

\ack{This research was supported by Basic Science Research Program through the National Research Foundation of Korea(NRF) funded by the Ministry of Education(2015-A419-0109).}

%%%%%%%%%%%%%%%%%%%%%%%%%%%%%%%%%

\appendix
\section{Metric in Kerr-AdS spacetime}

The metric are obtained from the line element of Kerr-AdS spacetime in equation (2.1) 
as $ds^2=g_{\mu\nu}dx^{\mu}dx^{\nu}$:
\begin{eqnarray}
g_{tt}&=&\frac{1}{\rho^2}(-\Delta_{r}+_Delta_{\theta}a^2\sin^2{\theta}), \ \ 
g_{t\phi}=g_{\phi t}=\frac{a\sin^2{\theta}}{\rho^2\Xi}
(\Delta_{r}-\Delta_{\theta}a^2\sin^2{\theta})\nonumber\\
g_{\phi\phi}&=&\frac{\sin^2{\theta}}{\rho^2\Xi^2}(\Delta_{\theta}^2 
-\Delta_{r}a^2\sin^2{\theta}), \ \ 
g_{rr}=\frac{\rho^2}{\Delta_{r}}, \ \ g_{\theta\theta}=\frac{\rho^2}{\Delta_{\theta}}.
\end{eqnarray}
The determinant of metric is given as 
\begin{eqnarray}
\sqrt{-g}=\sqrt{\rm{det}(-g_{\mu\nu})}=\frac{\rho^2\sin{\theta}}{\Xi}.
\end{eqnarray}

%%%%%%%%%%%%%%%%%%%%%%%%%%%%%%%%%%%%%%%%%%%%%%%%%%%%%%%
\section{Vierbein in Kerr-AdS spacetime}

The Vierbein in Kerr-AdS spacetime are obtained through the relation 
 $\hat{e}^{i}=b_{\mu}^{i}d{x}^{\mu}$ with the coordinate relation in equation (3.1):
\begin{eqnarray}
b^{\tilt}_{t}&=&\frac{\sqrt{\Delta_{r}}}{\rho}, \ \ b^{\tilt}_{\phi}=-\frac{\sqrt{\Delta_{r}}a\sin^2{\theta}}{\rho\Xi}, \nonumber\\ 
b^{\tilph}_{t}&=&-\frac{\sqrt{\Delta_{\theta}}a\sin{\theta}}{\rho}, \ \ 
b^{\tilph}_{\phi}=\frac{\sqrt{\Delta_{\theta}}\sin{\theta}(r^2+a^2)}{\rho\Xi}, \nonumber\\
b^{\tilr}_{r}&=&\frac{\rho}{\sqrt{\Delta_{r}}}, \ \ b^{\tilth}_{\theta}=\frac{\rho}{\sqrt{\Delta_{\theta}}}. 
 \end{eqnarray}
The inverse Vierbein are obtained by the relation 
$\eta_{ij}b_{\mu}^{i}b_{\nu}^{j}=g_{\mu\nu}, \ \rm{diag}(\eta_{\ij})=(-1,1,1,1,)$ as
\begin{eqnarray}
b_{\tilt}^{t}&=&\frac{r^2+a^2}{\rho\sqrt{\Delta_{r}}}, \ \ 
b_{\tilt}^{\phi}=\frac{a\Xi}{\rho\sqrt{\Delta_{r}}}, \nonumber\\ 
b_{\tilph}^{t}&=&\frac{a\sin{\theta}}{\rho\sqrt{\Delta_{\theta}}}, \ \ 
b_{\tilph}^{\phi}=\frac{\Xi}{\rho{\sqrt{\Delta_{\theta}}}\sin{\theta}}, \nonumber\\
b_{\tilr}^{r}&=&\frac{\sqrt{\Delta_{r}}}{\rho}, \ \ 
b^{\tilth}_{\theta}=\frac{\sqrt{\Delta_{\theta}}}{\rho}. 
 \end{eqnarray}

%%%%%%%%%%%%%%%%%%%%%%%%%%%%%%%%%%%%%%%%%%%%%%%%%%%%%%%
\section{Spin connection formula}

The definition of the spin connection $\omega^{ij, \mu}$ is obtained from the 
vierbein hypothesis:
\begin{equation}
\mathcal{D}_{\mu}b^k_{\nu}=D_{\mu}b^k_{\nu}-\Gamma^{\lambda}_{\nu\mu}b^k_{\lambda}
=\partial_{\mu}b^k_{\nu}+\omega^{k}_{j,\mu}b^j_{\nu}-\Gamma^{\lambda}_{\nu\mu}b^k_{\lambda}
=0,
\end{equation}
as 
\begin{equation}
\omega^{ij,k}=\Gamma^{\lambda}_{\nu\mu}-b_{k}^{\lambda}\partial_{\mu}b^k_{\nu}, 
\end{equation}
where $\Gamma^{\lambda}_{\nu\mu}$ denotes the Christoffel's 3-index symbol.
Using the identity $\gamma_k\gamma_{ij}=\eta_{ki}\gamma_j-\eta_{kj}\gamma_i+\gamma_{ijk}$, 
the spin connection term in the Dirac equation (3.2) is written as
\begin{equation}
\frac{1}{4}\gamma^{\mu}\omega^{ij}_{\mu}\gamma_{ij}=
\frac{1}{4}\omega^{ij}_{k}\gamma^{k}\gamma_{ij}=
\frac{1}{4}\omega^{ij}_{k}(-2\eta_{kj}\gamma_{i}+\gamma_{kij}).
\end{equation} 
The first term on the right-hand side in eq. (C.3) is written in a compact form as 
\begin{equation}
-\frac{1}{2}\omega^{ij}_{k}\eta_{kj}\gamma_{i}=
\frac{1}{4}\gamma^{\lambda}g^{\mu\nu}(\partial_{\lambda}g_{\mu\nu}-2\partial_{\mu}g_{\lambda\nu})
=\frac{\gamma^i}{2\sqrt{-g}}\partial_{\mu}(\sqrt{-g}b^{i\mu}),
\end{equation}
and the second term is written as
\begin{equation}
\frac{1}{4}\omega^{ij}_{k}\gamma_{kij}=\frac{1}{4}b^{\mu}_ib^{\nu}_j\partial_{\mu}b_{k\nu}\gamma_{kij}. 
\end{equation}
The sum of eqs. (C.4) and (C.5) is the spin connection formula in eq. (3.3).

\end{document}